\documentclass[10pt,a4paper,onecolumn]{article}
\usepackage{marginnote}
\usepackage{graphicx}
\usepackage[rgb]{xcolor}
\usepackage{authblk,etoolbox}
\usepackage{titlesec}
\usepackage{calc}
\usepackage{tikz}
\usepackage[pdfa]{hyperref}
\usepackage[numbers]{natbib}
\usepackage{hyperxmp}
\hypersetup{%
    unicode=true,
    pdfapart=3,
    pdfaconformance=B,
    pdftitle={STARRED: a two-channel deconvolution method with
Starlet regularization},
    pdfauthor={Kevin Michalewicz, Martin Millon, Frédéric Dux, Frédéric Courbin},
    pdfpublication={Journal of Open Source Software},
    pdfpublisher={Open Journals},
    pdfissn={2475-9066},
    pdfpubtype={journal},
    pdfvolumenum={},
    pdfissuenum={},
    pdfdoi={10.21105/joss.05340},  
    pdfcopyright={Copyright (c) 2023, Kevin Michalewicz, Martin Millon, Frédéric Dux, Frédéric Courbin},  
    pdflicenseurl={http://creativecommons.org/licenses/by/4.0/},
    colorlinks=true,
    linkcolor=[rgb]{0.0, 0.5, 1.0},
    citecolor=Blue,
    urlcolor=[rgb]{0.0, 0.5, 1.0},
    breaklinks=true
}
\immediate\pdfobj stream attr{/N 3} file{sRGB.icc}
\pdfcatalog{%
  /OutputIntents [
    <<
      /Type /OutputIntent
      /S /GTS_PDFA1
      /DestOutputProfile \the\pdflastobj\space 0 R
      /OutputConditionIdentifier (sRGB)
      /Info (sRGB)
    >>
  ]
}
\usepackage{caption}
\usepackage{orcidlink}
\usepackage{tcolorbox}
\usepackage{amssymb,amsmath}
\usepackage{ifxetex,ifluatex}
\usepackage{seqsplit}
\usepackage{xstring}

\usepackage{float}
\let\origfigure\figure
\let\endorigfigure\endfigure
\renewenvironment{figure}[1][2] {
    \expandafter\origfigure\expandafter[H]
} {
    \endorigfigure
}

\newlength{\cslhangindent}
\newlength{\csllabelwidth}
\newenvironment{CSLReferences}[2] 
 {
  \setlength{\parindent}{0pt}
  \ifodd #1 \everypar{\setlength{\hangindent}{\cslhangindent}}\ignorespaces\fi
  \ifnum #2 > 0
  \setlength{\parskip}{#2\baselineskip}
  \fi
 }%
 {}
\usepackage{calc}

\usepackage[top=3.5cm, bottom=3cm, right=1.5cm, left=1.0cm,
            headheight=2.2cm, reversemp, includemp, marginparwidth=4.5cm]{geometry}

\titleformat{\section}
  {\normalfont\sffamily\Large\bfseries}
  {}{0pt}{}
\titleformat{\subsection}
  {\normalfont\sffamily\large\bfseries}
  {}{0pt}{}
\titleformat{\subsubsection}
  {\normalfont\sffamily\bfseries}
  {}{0pt}{}
\titleformat*{\paragraph}
  {\sffamily\normalsize}

\usepackage{fancyhdr}
\makeatletter
\let\ps@plain\ps@fancy
\fancyheadoffset[L]{4.5cm}
\fancyfootoffset[L]{4.5cm}

\definecolor{linky}{rgb}{0.0, 0.5, 1.0}

\newtcolorbox{repobox}
   {colback=red, colframe=red!75!black,
     boxrule=0.5pt, arc=2pt, left=6pt, right=6pt, top=3pt, bottom=3pt}

\newcommand{\ExternalLink}{%
   \tikz[x=1.2ex, y=1.2ex, baseline=-0.05ex]{%
       \begin{scope}[x=1ex, y=1ex]
           \clip (-0.1,-0.1)
               --++ (-0, 1.2)
               --++ (0.6, 0)
               --++ (0, -0.6)
               --++ (0.6, 0)
               --++ (0, -1);
           \path[draw,
               line width = 0.5,
               rounded corners=0.5]
               (0,0) rectangle (1,1);
       \end{scope}
       \path[draw, line width = 0.5] (0.5, 0.5)
           -- (1, 1);
       \path[draw, line width = 0.5] (0.6, 1)
           -- (1, 1) -- (1, 0.6);
       }
   }

\patchcmd{\@maketitle}{center}{flushleft}{}{}
\patchcmd{\@maketitle}{center}{flushleft}{}{}
\patchcmd{\@maketitle}{\LARGE}{\LARGE\sffamily}{}{}  
\def\maketitle{{%
  
  \AB@maketitle}}
\renewcommand\AB@affilsepx{ \protect\Affilfont}
\renewcommand\AB@affilnote[1]{{\bfseries #1}\hspace{3pt}}
\renewcommand{\affil}[2][]%
   {\newaffiltrue\let\AB@blk@and\AB@pand
      \if\relax#1\relax\def\AB@note{\AB@thenote}\else\def\AB@note{#1}%
        \setcounter{Maxaffil}{0}\fi
        \begingroup
        \let\href=\href@Orig
        \let\protect\@unexpandable@protect
        \def\thanks{\protect\thanks}\def\footnote{\protect\footnote}%
        \@temptokena=\expandafter{\AB@authors}%
        {\def\\{\protect\\\protect\Affilfont}\xdef\AB@temp{#2}}%
         \xdef\AB@authors{\the\@temptokena\AB@las\AB@au@str
         \protect\\[\affilsep]\protect\Affilfont\AB@temp}%
         \gdef\AB@las{}\gdef\AB@au@str{}%
        {\def\\{, \ignorespaces}\xdef\AB@temp{#2}}%
        \@temptokena=\expandafter{\AB@affillist}%
        \xdef\AB@affillist{\the\@temptokena \AB@affilsep
          \AB@affilnote{\AB@note}\protect\Affilfont\AB@temp}%
      \endgroup
       \let\AB@affilsep\AB@affilsepx
}
\makeatother

\renewcommand\Affilfont{\sffamily\small\mdseries}
\ifnum 0\ifxetex 1\fi\ifluatex 1\fi=0 
  \usepackage[T1]{fontenc}
  \usepackage[utf8]{inputenc}

\else 
  \ifxetex
    \usepackage{mathspec}
    \usepackage{fontspec}

  \else
    \usepackage{fontspec}
  \fi
  \defaultfontfeatures{Scale=MatchLowercase}
  \defaultfontfeatures[\sffamily]{Ligatures=TeX}
  \defaultfontfeatures[\rmfamily]{Ligatures=TeX,Scale=1}

\fi
\IfFileExists{upquote.sty}{\usepackage{upquote}}{}
\usepackage{microtype}
\UseMicrotypeSet[protrusion]{basicmath} 

\PassOptionsToPackage{usenames,dvipsnames}{color} 
\ifLuaTeX
\usepackage[bidi=basic]{babel}
\else
\usepackage[bidi=default]{babel}
\fi
\babelprovide[main,import]{american}

\def\languageshorthands#1{}
\usepackage{longtable,booktabs,array}

\IfFileExists{parskip.sty}{%
\usepackage{parskip}
}{
\setlength{\parindent}{0pt}
\setlength{\parskip}{6pt plus 2pt minus 1pt}
}
\ifx\paragraph\undefined\else
\let\oldparagraph\paragraph
\renewcommand{\paragraph}[1]{\oldparagraph{#1}\mbox{}}
\fi
\ifx\subparagraph\undefined\else
\let\oldsubparagraph\subparagraph
\renewcommand{\subparagraph}[1]{\oldsubparagraph{#1}\mbox{}}
\fi
\ifLuaTeX
  \usepackage{selnolig}  
\fi

\usepackage{datetime}  
\newdateformat{subdate}{\twodigit{\THEDAY}~\monthname[\THEMONTH]~\THEYEAR}  

\title{STARRED: a two-channel deconvolution method with
Starlet regularization}

\author[1,2%
]{Kevin Michalewicz%
  \,\orcidlink{0000-0002-6584-2749}\,%
}
\author[1,3%
]{Martin Millon%
  \,\orcidlink{0000-0001-7051-497X}\,%
}
\author[1%
]{Frédéric Dux%
  \,\orcidlink{0000-0003-3358-4834}\,%
}
\author[1%
]{Frédéric Courbin%
  \,\orcidlink{0000-0003-0758-6510}\,%
}
\affil[1]{Institute of Physics, Laboratory of Astrophysics, École Polytechnique Fédérale de Lausanne (EPFL), Switzerland}
\affil[2]{Department of Mathematics, Imperial College London, London SW7 2AZ, United Kingdom}
\affil[3]{Kavli Institute for Particle Astrophysics and Cosmology and Department of Physics, Stanford University, Stanford, CA 94305, USA}
\date{\vspace{-2.5ex}}

\begin{document}
\maketitle

\marginpar{

  \begin{flushleft}
  \sffamily\small

  {\bfseries DOI:} \href{https://doi.org/10.21105/joss.05340}{\color{linky}{10.21105/joss.05340}}  

  \vspace{2mm}

  {\bfseries Software}
  \begin{itemize}
    \setlength\itemsep{0em}
    \item \href{https://github.com/openjournals/joss-reviews/issues/5340}{\color{linky}{Review}} \ExternalLink  
    \item \href{https://gitlab.com/cosmograil/starred}{\color{linky}{Repository}} \ExternalLink  
    \item \href{https://zenodo.org/record/7900798\#.ZFfIbXbMIq0}{\color{linky}{Archive}} \ExternalLink  
  \end{itemize}

  \vspace{2mm}

  \par\noindent\hrulefill\par

  \vspace{2mm}

  {\bfseries Editor:} \href{https://www.mpia.de/homes/momcheva/}{Ivelina Momcheva} \ExternalLink \orcidlink{0000-0003-1665-2073}
   \\
  \vspace{1mm}
    {\bfseries Reviewers:}
  \begin{itemize}
  \setlength\itemsep{0em}
    \item \href{https://github.com/ConnorStoneAstro}{@ConnorStoneAstro}
    \item \href{https://github.com/sibirrer}{@sibirrer}
    \end{itemize}
    \vspace{2mm}

  {\bfseries Submitted:} 10 March 2023 \\  
  {\bfseries Published:} 05 May 2023

  \vspace{2mm}
  {\bfseries License}\\
  Authors of papers retain copyright and release the work under a Creative Commons Attribution 4.0 International License (\href{https://creativecommons.org/licenses/by/4.0/}{\color{linky}{CC BY 4.0}}).

  \end{flushleft}
  
}

\hypertarget{summary}{%
\section{Summary}\label{summary}}

The spatial resolution of astronomical images is limited by atmospheric turbulence and diffraction in the telescope optics,
resulting in blurred images. This makes it difficult to accurately measure the brightness of blended objects because the contributions from adjacent objects are mixed in a time-variable manner due to changes in the atmospheric conditions. However, this effect can be corrected by characterizing the Point Spread Function (PSF), which describes how a point source is blurred on a detector. This function can be estimated from the stars in the field of view, which provides a natural sampling of the PSF across the entire field of view.

Once the PSF is estimated, it can be removed from the data through the so-called deconvolution process, leading to images of improved spatial resolution. The deconvolution operation is an ill-posed inverse problem due to noise and pixelization of the data. To solve this problem, regularization is necessary to guarantee the robustness of the solution. Regularization can take the form of a \textit{sparse} prior, meaning that the recovered solution can be represented with only a few basis eigenvectors.

\texttt{STARRED} is a Python package developed in the context of the \href{www.cosmograil.org}{COSMOGRAIL} collaboration and applies to a vast variety of astronomical problems. It proposes to use an isotropic wavelet basis, called Starlets (\protect\hyperlink{ref-Starck:2015}{Starck et al., 2015}), to regularize the solution of the deconvolution problem. This family of wavelets has been shown to be well-suited to represent astronomical objects. \texttt{STARRED} provides two modules to first reconstruct the PSF, and then perform the deconvolution. It is based on two key concepts: i) the image is reconstructed in two separate \textit{channels}, one for the point sources and one for the extended sources, and ii) the code relies on the deliberate choice of not completely removing the effect of the PSF, but rather bringing the image to a higher resolution with a known Gaussian PSF. This last point allows us to suppress the deconvolution artifacts, which occur when attempting to deconvolve to an infinite resolution, as most of other techniques do. Finally, \texttt{STARRED} uses JAX automatic differentiation to ensure gradient-informed optimization of this high-dimension optimization problem (\protect\hyperlink{ref-Bradbury:2018}{Bradbury et al., 2018}).

\hypertarget{statement-of-need}{%
\section{Statement of need}\label{statement-of-need}}

\begin{figure}[h]
    \centering
    \includegraphics[width=\textwidth]{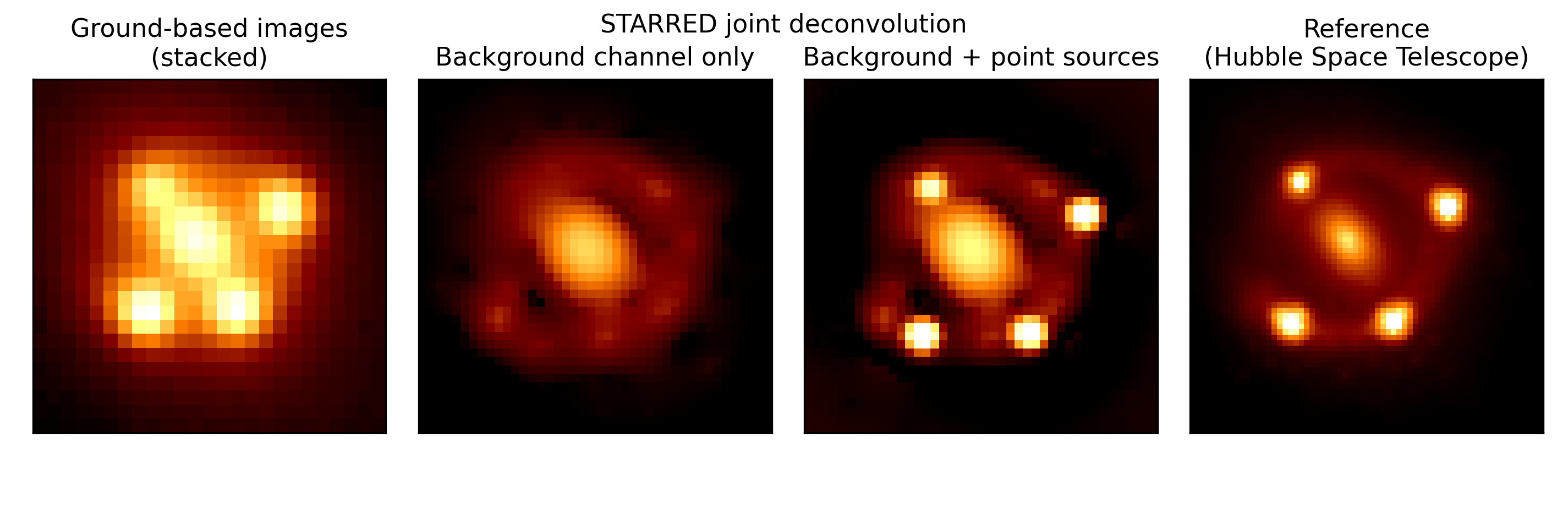}
    \caption{Multi-epoch deconvolution of the DES2038-4008 lensed quasar, SDSS-r filter. From left to right, the panels show: a stacked image of all 31 epochs used in the deconvolution (32x32 pixels); \texttt{STARRED} multi-epoch deconvolution showing only the background (second panel, 64x64 pixels) and also including the point sources (third panel, 64x64 pixels) with a runtime of 17 seconds (GPU: RTX 3060); a deep, high-resolution observation (\protect\hyperlink{ref-Shajib:2018}{Shajib et al., 2018}) from the Hubble Space Telescope, F160W filter (Credit: TDCOSMO Collaboration, GO-15320, PI: Treu).}
    \label{deconv}
\end{figure}

Image deconvolution is a widespread problem, not only in astronomy, but also in other scientific fields such as medical imaging and microscopy. In astronomy, \texttt{STARRED} can be applied to a wide range of cases, such as extracting light curves from lensed quasars, extracting light curves of Cepeheids in crowded fields, deblending supernovae and their host galaxies, among many others. The software is not limited to single images, but can also process time series data. This is particularly useful in the context of large time-domain sky surveys like Rubin-LSST\footnote{\texttt{https://www.lsst.org/}}. \texttt{STARRED} can jointly deconvolve multiple observation epochs based on their respective estimated PSFs, thereby removing epoch-to-epoch PSF variations before obtaining photometric measurements. An example of such multi-epoch deconvolution is shown in Figure \ref{deconv}. Here, the position of the point sources and the extended source channel are constrained across all epochs, whereas the amplitude of the point sources can vary thoughout the time series.

The most popular early image deconvolution methods include Högbom (\protect\hyperlink{ref-Hogbom:1974}{1974}) and Lucy \& Walsh (\protect\hyperlink{ref-Lucy:2003}{2003}). The latter, the Richardson-Lucy algorithm, was updated to decompose images into point source and extended source channels (\protect\hyperlink{ref-Becker:2003}{Becker et al., 2003}). An extension to more than two channels was proposed in Bontekoe et al. (\protect\hyperlink{ref-Bontekoe:1994}{1994}). These works have been further developed and enhanced under Bayesian frameworks (\protect\hyperlink{ref-Selig:2015}{Selig \& Enßlin, 2015}) and even using neural networks with a 1-channel approach (\protect\hyperlink{ref-Akhaury:2022}{Akhaury et al., 2022}; \protect\hyperlink{ref-Sureau:2020}{Sureau et al., 2020}).

What all these algorithms have in common is that they intend to fully correct for the PSF, which leads to artifacts due to the poor representation of frequencies aliased outside the frequency domain allowed by the pixel size of the deconvolved image. In other words, should this effect be completely eliminated, then it would be necessary to sample the deconvolved image at infinitely small intervals. This drawback is overcome with MCS (\protect\hyperlink{ref-Magain:1998}{Magain et al., 1998}) since a narrower version of the original PSF is generated to satisfy Nyquist-Shannon sampling theorem, and thus its outputs are not prone to artifacts.

\texttt{STARRED} is based on the principles of the Fortran code MCS and Firedec (\protect\hyperlink{ref-Cantale:2016}{Cantale et al., 2016}). However, \texttt{STARRED} has several important improvements over these two methods: i) it uses a family of isotropic wavelets well-suited to represent astronomical objects, Starlets, ii) it is auto-differentiable, facilitating the use of optimizers based on gradient descent, iii) it is faster than previous codes (220 CPU seconds versus 260 with MCS for the case of Figure \ref{deconv}) and GPU scalable, and iv) it can handle time series by processing all data together in a multi-epoch deconvolution. The use of a second channel for the point sources allows us to explicitly account for the highest frequencies, which cannot be sparsely represented with Starlets. 

The scalability, speed of the code and its ability to handle time series are especially relevant for processing the large volume of data expected from future large sky time-domain surveys. In addition, \texttt{STARRED} is much more robust to local minima than MCS or Firedec since it uses gradient descent with momentum algorithms and because of a well-justified regularization, \textit{i.e.}, Starlets, where galaxies are known to be sparse.

\hypertarget{Functionality}{%
\section{Functionality}\label{implementation}}
The main functionality of \texttt{STARRED} is built around deconvolution and PSF generation classes. The \texttt{starred.psf} module receives as input several stars of the field of interest and optionally the noise maps of the observations. It is capable of computing the convolution kernel that relates the original image to an image with a known Gaussian PSF with a Full Width at Half Maximum (FWHM) of 2 pixels. We call it \textit{the narrow PSF}. It is expressed as a sum of an analytic function with circular symmetry and a Starlet regularized grid of pixels.

The \texttt{starred.deconvolution} module features all the necessary classes to deconvolve an image or a time series. It is necessary to provide the observations, the retrieved \textit{narrow} PSF and, optionally, the noise maps. This step decomposes the deconvolved image into two channels of point sources and extended sources. The latter are represented by a Starlet regularized grid of pixels. During this process, the image is decomposed into several scales, each of which captures particular frequency features. It is then reconstructed from only the highest coefficients of each scale, which contain the signal but not the noise. This procedure, referred as \textit{soft thresholding}, allows us to remove the noise from both the reconstructed PSF and the deconvolved image. 

\texttt{STARRED} is implemented in JAX for a fast computation of all the derivatives of the free parameters with automatic differentiation. It relies on several gradient descent algorithms from the \texttt{optax} and \texttt{scipy} Python packages.

When processing time series, this module allows us to perform a joint deconvolution of all the images, benefiting from the combined signal-to-noise ratio of all frames but allowing to measure the photometry of the point sources in each individual image. This module can therefore extract the light curves of blended point sources even superposed on a background of any arbitrary shape. 

Finally, the \texttt{starred.plots} module can be used to generate web-friendly images with the appropriate cuts.

\hypertarget{acknowledgements}{%
\section{Acknowledgements}\label{acknowledgements}}

We acknowledge the support of the Swiss National Science Foundation (SNSF) and the European Research Council (ERC) under the European Union’s Horizon 2020 research and innovation programme (COSMICLENS: grant agreement No 787886). Kevin Michalewicz acknowledges support from the President's PhD Scholarship at Imperial College London.

\hypertarget{references}{%
\section*{References}\label{references}}
\addcontentsline{toc}{section}{References}

\hypertarget{refs}{}
\begin{CSLReferences}{1}{0}
\leavevmode\vadjust pre{\hypertarget{ref-Akhaury:2022}{}}%
Akhaury, U., Starck, J.-L., Jablonka, P., Courbin, F., \& Michalewicz, K. (2022). Deep
learning-based galaxy image deconvolution.  \emph{Frontiers in Astronomy and Space Sciences},
\emph{9}. \url{https://doi.org/10.3389/fspas.2022.1001043}

\leavevmode\vadjust pre{\hypertarget{ref-Becker:2003}{}}%
Becker, T., Fabrika, S., \& Roth, M. M. (2003). Crowded field 3D spectroscopy.  \emph{Astronomische
Nachrichten},
\emph{325}, 155–158. \url{ https://doi.org/10.1002/asna.200310198}

\leavevmode\vadjust pre{\hypertarget{ref-Bontekoe:1994}{}}%
Bontekoe, T. R., Koper, E., \& Kester, D. J. M. (1994). Pyramid maximum entropy images of
IRAS survey data. \emph{Astronomy and Astrophysics},
\emph{284}, 1037–1053.

\leavevmode\vadjust pre{\hypertarget{ref-Bradbury:2018}{}}%
Bradbury, J., Frostig, R., Hawkins, P., Johnson, M. J., Leary, C., Maclaurin, D., Necula, G.,
Paszke, A., VanderPlas, J., Wanderman-Milne, S., \& Zhang, Q. (2018). JAX: Composable
transformations of Python+NumPy programs (Version 0.3.13). \url{http://github.com/google/jax}

\leavevmode\vadjust pre{\hypertarget{ref-Cantale:2016}{}}%
Cantale, N., Courbin, F., Tewes, M., Jablonka, P., \& Meylan, G. (2016). Firedec: A two-channel finite-resolution image deconvolution algorithm. \emph{Astronomy and Astrophysics},
\emph{589}, A81. \url{https://doi.org/10.1051/0004-6361/201424003}

\leavevmode\vadjust pre{\hypertarget{ref-Hogbom:1974}{}}%
Högbom, J. A. (1974). Aperture Synthesis with a Non-Regular Distribution of Interferometer
Baselines. \emph{Astronomy \& Astrophysics Supplement},
\emph{15}, 417.

\leavevmode\vadjust pre{\hypertarget{ref-Lucy:2003}{}}%
Lucy, L. B., \& Walsh, J. R. (2003). Iterative techniques for the decomposition of long-slit
spectra. \emph{The Astronomical Journal},
\emph{125}(4), 2266. \url{https://doi.org/10.1086/368144}

\leavevmode\vadjust pre{\hypertarget{ref-Magain:1998}{}}%
Magain, P., Courbin, F., \& Sohy, S. (1998). Deconvolution with correct sampling. \emph{The
Astrophysical Journal},
\emph{494}(1), 472–477. \url{https://doi.org/10.1086/305187}

\leavevmode\vadjust pre{\hypertarget{ref-Selig:2015}{}}%
Selig, M., \& Enßlin, T. A. (2015). Denoising, deconvolving, and decomposing photon observations. \emph{Astronomy and Astrophysics},
\emph{574}, A74. \url{https://doi.org/10.1051/0004-6361/201323006}

\leavevmode\vadjust pre{\hypertarget{ref-Shajib:2018}{}}%
Shajib, A. J., Birrer, S., Treu, T., Auger, M. W., Agnello, A., Anguita, T., Buckley-Geer, E. J., Chan, J. H. H., Collett, T. E., Courbin, F., Fassnacht, C. D., Frieman, J., Kayo, I., Lemon, C., Lin, H., Marshall, P. J., McMahon, R., More, A., Morgan, N. D., ... Walker, A. R. (2018). Is every strong lens model unhappy in its own way? Uniform modelling of a sample of 13 quadruply+ imaged quasars. \emph{Monthly Notices of the Royal Astronomical
Society},
\emph{483}(4), 5649–5671. \url{ https://doi.org/10.1093/mnras/sty3397}

\leavevmode\vadjust pre{\hypertarget{ref-Starck:2015}{}}%
Starck, J.-L., Murtagh, F., \& Bertero, M. (2015). Starlet transform in astronomical data processing. In O. Scherzer (Ed.), \emph{Handbook of mathematical methods in imaging} (pp. 2053–2098). Springer New York. \url{. https://doi.org/10.1007/978-1-4939-0790-8_34}

\leavevmode\vadjust pre{\hypertarget{ref-Sureau:2020}{}}%
Sureau, F., Lechat, A., \& Starck, J.-L. (2020). Deep learning for a space-variant deconvolution in galaxy surveys. \emph{Astronomy and Astrophysics},
\emph{641}, A67. \url{https://doi.org/10.1051/0004-6361/201937039}

\end{CSLReferences}

\end{document}